\documentclass[12pt,preprint]{aastex}

\shorttitle{Temperature Variations in the Orion Nebula}
\shortauthors{Rubin et al.}

\begin{document}

\title{Temperature Variations from HST Spectroscopy of the
Orion Nebula}

\author{R.~H.~Rubin\altaffilmark{1,2}, 
P.~G.~Martin\altaffilmark{3},
R.~J.~Dufour\altaffilmark{4}, 
G.~J.~Ferland\altaffilmark{5}, 
K.~P.~M.~Blagrave\altaffilmark{3},\break
X.-W.\ Liu\altaffilmark{6},  
J.~F.~ Nguyen\altaffilmark{1} 
and
J.~A.\ Baldwin\altaffilmark{7}}
\altaffiltext{1}{NASA/Ames Research Centre, Moffett Field, CA 94035-1000, USA}
\altaffiltext{2}{Orion Enterprises, M.S. 245-6, Moffett Field, CA 94035-1000, 
USA}
\altaffiltext{3}{Canadian Institute for Theoretical Astrophysics, University 
of Toronto, Toronto, ON M5S~3H8 Canada}
\altaffiltext{4}{Physics \& Astronomy Department, Rice University, MS 61, 
Houston, TX 77005-1892, USA}
\altaffiltext{5}{Physics \& Astronomy Department, University of Kentucky,  
Lexington, KY 40506-0055, USA}
\altaffiltext{6}{Physics \& Astronomy Department, University College London, Gower 
Street, London, UK WC1E~6B}
\altaffiltext{7}{Physics \& Astronomy Department, Michigan State University, 
East Lansing, MI 48824-1116, USA}

\begin{abstract}
We present HST/STIS long-slit spectroscopy of NGC~1976.
Our goal is to measure the intrinsic line ratio [\ion{O}{3}] 
4364/5008 and thereby evaluate the
electron temperature ($T_e$) and the fractional mean-square $T_e$ variation 
($t_A^2$) {\it across the nebula}.
We also measure the intrinsic line ratio [\ion{N}{2}] 
5756/6585 in order to estimate $T_e$ and $t_A^2$ in the N$^+$ region. 
The interpretation of the [\ion{N}{2}] data is not as clear cut as the
[\ion{O}{3}] data because of a higher sensitivity to knowledge of the electron
density as well as a possible contribution 
to the [\ion{N}{2}] 5756 emission by recombination (and cascading).
We present results from binning the data along the various slits into tiles 
that are 0.5$''$ square (matching the slit width).
The average [\ion{O}{3}] temperature for our four HST/STIS
slits varies from 7678~K to 8358~K; $t_A^2$ varies from 0.00682 to 
at most 0.0176.
For our preferred solution, 
the average [\ion{N}{2}] temperature for each of the four 
slits varies from 9133~K to 10232~K; $t_A^2$ varies from 0.00584 to 0.0175.
The measurements of $T_e$ reported here are an average
along each line of sight.
Therefore, despite finding remarkably low $t_A^2$, we cannot rule out 
significantly larger temperature fluctuations along the line of sight.
The result that the average [\ion{N}{2}] $T_e$ exceeds the
average [\ion{O}{3}] $T_e$ confirms what has been 
previously found for Orion and what is expected on theoretical grounds.
Observations of the proplyd P159-350 indicate: large local extinction
associated; ionization stratification consistent with
external ionization by  $\theta^1$ Ori C; and indirectly, 
evidence of high electron density. 

\end{abstract}

\vskip-0.2truein

\keywords{ISM: abundances --- ISM: atoms --- ISM: \ion{H}{2} regions --- 
ISM: individual: NGC~1976 --- ISM: individual: proplyd P159-350}

\section{INTRODUCTION}

     Most observational tests of the chemical evolution of the universe
rest on emission line objects; these define the endpoints of stellar
evolution and probe the current state of the interstellar medium (ISM).
Gaseous nebulae (\ion{H}{2} regions \& Planetary Nebulae (PNs)) 
are laboratories for understanding physical 
processes in all emission-line sources, and probes for stellar, 
galactic, and primordial nucleosynthesis. 

	There is a fundamental issue that continues to be problematic~--
the discrepancy between heavy element
abundances inferred from emission lines that are collisionally excited 
(CELs)
compared with those 
due to recombination/cascading, the so-called ``recombination lines''.
Studies of PNs contrasting recombination and collisional abundances 
(Liu et~al.\ 1995, Kwitter \& Henry 1998)
often find differences exceeding a factor of two. 
For NGC~7009, Liu et~al.\ (1995) found 
that the recombination C, N, and O abundances are a factor of $\sim$5
larger than the corresponding collisional abundances,
which was found to be the case also for neon (Luo, Liu \& Barlow 2001).
For NGC~6153, Liu et~al.\ (2000) found
that C$^{++}$/H$^+$, 
N$^{++}$/H$^+$, 
O$^{++}$/H$^+$, and
Ne$^{++}$/H$^+$ ratios derived from optical recombination lines
are all a factor of $\sim$10 higher than the corresponding values
deduced from CELs.
The discrepancy is even larger,  
a factor of $\sim$20,
for the Galactic bulge PN M~1-42
(Liu et~al.\ 2001) and Hf~2-2, which has the most extreme 
abundance difference to date, a factor of 84 (Liu 2002).

	Studies of \ion{H}{2}  regions 
find similar behavior although the differences are generally smaller
than is the case for PNs.
For the Orion Nebula, Esteban et~al.\ (1998) derived a factor of 1.5
larger for O$^{++}$/H$^+$ from recombination lines than from CELs.
Using several ions for a study of M8, Esteban et~al.\ (1999) 
found again that the ionic abundances are always somewhat lower
as derived from collisional forbidden lines than those from
corresponding recombination lines.  
The mean abundance difference is a factor of $\sim$2.
Recently Tsamis et~al.\ (2002) presented observations of 
5 more \ion{H}{2}  regions~--
M~17 and NGC~3576 as well as the Magellanic Cloud 
\ion{H}{2}  regions 30 Doradus, LMC~N11B and SMC N66.
They found that the disparity was always in the same direction.
For four of their objects, the O$^{++}$/H$^+$ abundance
from \ion{O}{2} recombination lines exceeded the corresponding
value inferred from the nebular [\ion{O}{3}] CELs by factors
ranging from 1.8~-- 2.7, while the factor was $\sim$5 for 
LMC~N11B.   
According to the limited statistics, apparently \ion{H}{2} regions 
exhibit less of an abundance dichotomy than some of the PNs.

	Most of the efforts to explain the abundance 
puzzle between collisional and recombination values have attempted
to do so by examining electron temperature ($T_e$) variations in the plasma.
Because $T_e$ is not expected to vary dramatically within the (hydrogen)
ionized region of a given nebula (e.g., Harrington et~al.\ 1982; Baldwin 
et~al.\ 1991; Rubin et~al.\ 1991), 
a method developed by Peimbert (1967) has been extensively used.
He developed a useful formalism 
by expressing the volume emissivity for a given spectral line
in a Taylor series expansion around an 
average temperature defined such that the first order term vanishes.  
Since the [fractional]
mean-square $T_e$ variation (called $t^2$) is expected to be small, 
only terms to second order need be retained.
The resulting influence on elemental abundance determinations began with
the work of Peimbert (1967), Rubin (1969), 
and Peimbert \& Costero (1969).
There is a vast literature now that measures the discrepancy between 
collisional and recombination abundances in terms of $t^2$.

The studies of \ion{H}{2}  regions mentioned above
also derived the value of $t^2$ 
that forces the two abundance techniques to yield the same result.
For the Orion Nebula, Esteban et~al.\ (1998) 
required $t^2$~$\sim$ 0.024 to reconcile the difference,
while for M8, Esteban et~al.\ (1999) found an even
larger $t^2$~= 0.032 was needed on average for the various
ions in common.
Tsamis et~al.\ (2002) found for 30~Doradus
a similar value, $t^2$~$\sim$ 0.03.
For the drastically different O$^{++}$/H$^+$ abundances
found for NGC~7009, 
Liu et~al.\ (1995) discussed that it would be
necessary to invoke $t^2$~$\sim$ 0.1, 
which would then force 
agreement 
close to the higher recombination value~--
a value more than 2.5 times larger than the solar
O/H of 7.41$\times$10$^{-4}$ (Grevesse \& Sauval 1998).
Such a large $t^2$ is not at all predicted by current theory/models
(e.g., Kingdon \& Ferland 1998).

	The current unsettled situation
has led to efforts to broaden the study to include other 
variables besides $T_e$ to 
analyse the effects upon
abundance determinations.
One promising avenue is to examine abundance derivation
 considering density variations,
abundance variations, and $T_e$ variations in combination.
Liu et~al.\ (2000) took this approach in their investigation of
the PN NGC~6153 with a two-phase empirical model.
P\'equignot et~al.\ (2002) continued the study using
photoionization models including two components with different 
heavy element abundances.

	This is an extension of an earlier paper (Rubin et~al.\ 2002~-- 
Paper~I) that used
HST STIS and WFPC2 observations to study the variation of $T_e$ in NGC~7009
as determined from the [\ion{O}{3}] (4364/5008) flux ratio.
Very low values for $t^2$ in the plane of the sky 
($t_A^2$) were found (always $\la$0.01).
In this paper, we focus solely on $T_e$ variations with the purpose 
to determine from the observational data 
the magnitude of $t_A^2$ for the Orion Nebula.
In section 2, we present the {\it HST} observations
and data reduction procedures.
Section 3 includes a discussion and analysis of extinction.
We determine the electron temperature distributions in section 4.
In section 5, we analyse the $T_e$ distributions 
in terms of average temperatures and fractional mean-square 
temperature variations in the plane of the sky.
Section 6 provides a discussion and conclusions.

\section{{\it HST} OBSERVATIONS AND DATA REDUCTION}

	The observations of NGC~1976 described here were taken 
as part of our HST Cycle 7 program GO-7514.
We observed with 4 different STIS long-slits: slits 1, 2, 4, and 5.
These are shown in Figure~1.
Our slit positions are chosen to cross several features, including an
Herbig-Haro (HH) object, a proplyd,  and the Orion Bar. 
Slit 1 passes through
the position 1SW which we observed with FOS and GHRS in Cycle 5
(see Figure~1 in Rubin et~al.\ 1997).
The centre of 1SW is at 
$\alpha$, $\delta$ = $5^{\rm h}35^{\rm m}14\fs71$, 
$-5^{\rm o}$23\arcmin41\farcs5
(all positions are equinox J2000), 
18.5$''$ S and 26.2$''$ W of $\theta^1$ Ori C. 
This slit was also chosen to pass through proplyd P159-350 
(O'Dell \& Wen 1994).
Slit 2 passes through position x2 observed with FOS in
Cycle 5 
(see Figure~1 in Rubin et~al.\ 1997).
x2 is located in one of the most prominent arcs in
the [\ion{O}{3}] WFPC2 images. 
Slit~2 is parallel to slit 1, which allowed a substantial saving in 
overhead (OH).
The position angle (PA)~= 114.555$^{\rm o}$;
the distance between slit centres is 18.1$''$.
Note that the bottom of slit~1 and the top of slit~2
have a separation of only 0.211$''$.
The displacement between these two slits in the direction along
the slit  is 18.056$''$. 

	Slit 4 crosses the Orion Bar and is oriented to point toward 
$\theta^1$~Ori~C (the dominant exciting star), 
which also places it essentially orthogonal to the Bar.   
The southern tip passes through the Herbig-Haro object HH~203. 
Slit 5 passes through a very bright,
sharply-defined ``rim" of the Bar where a positional
bifurcation  begins. 
Slit 5 is parallel to slit 4 in order to take advantage of the same
OH savings as is the case for slits 1 and  2. 
The PA~= 139.068$^{\rm o}$;
distance between centres of slits 4 and  5 is $\sim$32$''$. 

	Observations were made as follows:
data for slits 1 and 2 are from Visit 2 (1998 December 7 UT), 
Visit 52 (2000 December 7)
and Visit 72 (2001 December 16);
data for slits 4 and 5 are from Visit 5 (1998 December 22).
All our data described here were acquired using the STIS/CCD with a 52$''$ slit
length and 0.5$''$ slit width.
Each visit comprises 2 orbits.
Spectra were taken with gratings G430M (with 
wavelength settings: 3680, 4451, 4961) 
and G750M (settings: 5734, 6581).  
Each exposure was done in accumulation mode and at least
two spectra were taken at each setting in order to cosmic-ray (CR) clean.
The data sets we processed were those obtained after sufficient time elapsed 
from the observation dates in order that ``best reference files" would
be stable/finalized.  We requested On-the-Fly Calibration for science files
and Best Reference Files.
After retrieving the data sets, we then co-added and cosmic ray 
cleaned images using standard packages 
in IRAF.\footnote{IRAF is distributed by NOAO, which is operated by AURA,
under cooperative agreement with NSF.}
Calibrations to produce 2-dimensional (2D) rectified images 
were then carried out.
 From these, we singled out specific emission lines for further investigation.
Data for the [\ion{O}{3}] 4364~\AA\ line as well as H$\gamma$ 4342~\AA\
were contained in the G430M/4451 grating setting;
for the [\ion{O}{3}] 5008, 4960~\AA\ lines and H$\beta$ 4863~\AA\
lines, G430M/4961 was used;
for [\ion{N}{2}] 5756~\AA, G750M/5734 was used;
for H$\alpha$ 6565~\AA\ and the
[\ion{N}{2}] 6550, 6585~\AA\ lines, G750M/6581 was used.
All wavelengths in this paper are vacuum rest wavelengths.

For G430M, the dispersion is 0.28~\AA\ per pixel 
for a point source and the plate scale 0.05 arcsec/pixel
(Leitherer et~al.\ 2001, Chapter~13).
For a uniformly filled slit
with width 0.5$''$, a degradation in
resolving power by a factor of 10 is expected
to a spectral resolution of 2.8~\AA.  
For G750M, the dispersion is 0.56~\AA\ per pixel 
for a point source and the plate scale 0.05 arcsec/pixel.
For a uniformly filled slit
with width 0.5$''$, the spectral resolution would be 5.6~\AA.

	According to Paul Goudfrooij (private communication),
the STIS CCD absolute flux calibration that is performed in the pipeline
follows methods described 
in Instrument Science
Report (ISR) 97-14 (Bohlin, Collins \& Gonnella 1998)
available through the STIS web site at
www.stsci.edu/hst/stis/documents/isrs.
That report considers only the L (low-resolution) modes specifically,
but the flux calibration for the M (medium-resolution) modes is done the same way,
i.e., by comparing the observed spectrum of a primary standard with a
pure-hydrogen white dwarf model. 
The calibration observations were made
with the 52$''$$\times$2$''$ slit;
thus relative transmission corrections are necessary
to derive the absolute fluxes for the other slits (such as the 
52$''$$\times$0.5$''$ slit we used).
This procedure is documented in ISR 98-20 (Bohlin \& Hartig 1998).

	For the analysis presented here, we were interested mainly in
the distribution of line flux along the slit spatial direction.
This was accomplished with the IRAF routine {\it blkavg} in conjunction
with specialized software tools that we developed ourselves.
	Even after applying the standard CR rejection 
there still remain many bad pixels due to CRs and/or hot pixels.
There is considerable danger that including these can corrupt 
the flux values we seek.  The program developed to eliminate these
remaining bad pixels is called PIXHUNTER, which has been described briefly 
in an earlier paper (see Appendix~A of Paper~I).
Once the columns containing the line have been cleaned for 
deviant pixels, we are ready to subtract an 
equivalent spectral range of continuum.
We do this by using IRAF functions, including {\it blkavg},
to operate upon the appropriate sections of cleaned continuum. 
The 1-dimensional (1D) distribution of line flux versus spatial coordinate
for the various emission lines of interest is what we need
for our subsequent physical analysis.

	We note that there is excellent agreement with 
a cross check of the 1D results of flux versus spatial direction
by comparing with 1D results of flux versus wavelength for a corresponding
spatial sample.
The latter were measured with the {\it splot} package.
With this, the underlying continuum is fitted and the integrated
line flux determined with the $e$-option (area under the line profile), which
was preferable to fitting with a Gaussian profile.
Because of the spectral impurity introduced by the
relatively wide slit used, the line profiles have flatter tops and less 
extended bases (i.e., they are more ``trapezoidal'') 
than the Gaussian fits.  It is also apparent that
the Gaussian fit is overestimating the line flux.  

	Both the [\ion{O}{3}] 5008 and 4960~\AA\ lines were 
observed simultaneously with the\break
G430M/4961 grating setting.
Because both transitions arise from the same upper level,
the intrinsic flux ratio depends only on the transition probabilities
(A-values) and wavelengths.
As reported previously for similar STIS observations of
the planetary nebula NGC~7009 
obtained under program GO-8114 (PI RR)
(Paper~I), 
what we found was a surprising variation in the 
F(5008)/F(4960) ratio with position along the slit.
This amounts to a variation in the ratio of roughly 3.0$\pm$0.1.
Furthermore, it appears more-or-less 
periodic with an $\sim$3.5$''$ cycle.
As described in Paper~I,
according to Ted Gull (private communication),
this is an instrumental effect and is a ratio of two fringe patterns. 
The source of the problem is a thin blocker filter that had to be matched 
with each grating and the best (and only) location that it could be placed was 
above the grating in a stable mounting. 
	To attempt to do anything about fringing would probably require a
dedicated HST/STIS calibration program.  
If there were fringing in the F(5008)/F(4364) ratio at the same level
as for the F(5008)/F(4960) ratio, the $\pm$ 3.3\% error would 
result in only a minor $T_e$ error, e.g., $\pm$ 100~K at $T_e$~= 10$^4$~K
(see \S4.1).

\section{EXTINCTION and REDDENING CORRECTION}

	Before deriving the $T_e$ distribution from the
STIS data, we first correct for extinction.
This is calculated by comparing the observed 
F(H$\alpha$)/F(H$\beta$) ratio with the theoretical ratio 
I(H$\alpha$)/I(H$\beta$).
We use a value of 2.88 assuming $T_e$~= 8500~K and $N_e$~= 
5000~cm$^{-3}$, Case B (Storey \& Hummer 1995).\footnote{For conditions
applicable to NGC~1976, we find using an online program 
(see Storey \& Hummer 1995) that the 
I(H$\alpha$)/I(H$\beta$) ratio used here will depart by less than 
3\% over the range  7500 $\le$ $T_e$ $\le$ 12500~K and
10$^3$ $\le$ $N_e$ $\le$ 10$^5$~cm$^{-3}$.}
The extinction correction is done in terms of
$c$(H$\beta$),
given by the relationship
\begin{equation} 
log [F(\lambda)/F(H\beta)] = log [I(\lambda)/I(H\beta)] - f(\lambda)~ c(H\beta),
\end{equation}
\noindent
where $f(\lambda$) is the extinction curve.
For the H$\gamma$ (4342), 4364, 4863, 4960, 5008, 5756, 6550, 6565, and 
6585 \AA\ lines, the respective values for $f(\lambda$) are 
0.0856, 0.082, 0, $-$0.0153,  $-$0.0225, $-$0.1237, $-$0.2185,
$-$0.2202, and $-$0.2226,  (Martin et al. 
1996).
This leads to, 
\begin{equation} 
c(H\beta) = 4.541~ log [F(H\alpha)/F(H\beta)] -  2.086 .
\end{equation}

	The correction for extinction/reddening from observed
to intrinsic flux for the 4364 and 5008 lines is then given by,
\begin{equation} 
I(4364) = F(4364)~ 10^{1.082~ c(H\beta)}~~; 
~~~I(5008) = F(5008)~ 10^{0.9775~ c(H\beta)}~~~~  ,
\end{equation}
\noindent
and for the 5756 and 6585 lines by,
\begin{equation} 
I(5756) = F(5756)~ 10^{0.8763~ c(H\beta)}~~; 
~~~I(6585) = F(6585)~ 10^{0.7774~ c(H\beta)}~~~~  .
\end{equation}
\noindent

	For the STIS data, we binned the pixels along the slit into 
tiles  that are 0.5$''$ square (matching the slit width).
The STIS fiducial bars are excluded.
This produced $c$(H$\beta$) results for from 79 to 96 tiles, depending on
the slit/visit  observed, which are the same set used later for the $T_e$ analysis.
The distributions of $c$(H$\beta$) along the two  slits (slit~1 and slit~2) 
observed in 3 separate visits (V2, V52, and V72) are remarkably similar.
Furthermore, the distributions match well for the regions of slit~1 and slit~2 
that are adjacent to each other.  Recall that the separation
between the bottom of slit~1 and the top of slit~2 is a mere 0.211$''$
and that slit~2 is shifted 18.056$''$ relative to slit~1 along the slit
spatial direction.
The average $c$(H$\beta$) values
without regard to any weighting for brightness are:
for slit~1: 
0.680, 0.699, and 0.701
respectively for the three visits;
for slit~2:
0.762, 0.798, and 0.766.
For slit~4, it is 0.566
and for slit~5, it is 0.611.
We comment further on slit~1, which covers the proplyd P159-350
and 1SW.
Starting at the SE end of slit~1, 
the values are roughly flat at $\sim$0.85.
There is a spike to 1.43 in V2 at the tile containing most of the proplyd. 
The behavior is similar for V72 with 
a spike to 1.33 at the tile containing most of the proplyd. 
The adjacent nebula has for both visits  
a $c$(H$\beta$) value of $\sim$0.9.
Unfortunately, because P159-350 was centred in the East
fiducial bar on V52, we cannot reliably measure
$c$(H$\beta$) there.
Where slit~1 crosses 1SW, $c$(H$\beta$) is $\sim$0.6, which 
agrees well with our previous spectroscopic value 0.605 at our 
position observed with the Faint Object Spectrograph (FOS-1SW, Rubin et~al.\ 1998).
At the end of slit~1 toward the NW, 
$c$(H$\beta$) has decreased to $\sim$0.4.

	We examined the vicinity of P159-350 in more detail using the 
original resolution of 1 pixel (0.05$''$) for slit~1 in the spatial direction.  
The increase in flux and thus signal-to-noise (S/N) in the vicinity 
of the proplyd permits meaningful analysis here.
For V2, we find that the peak observed flux occurs in pixel 311 for
the H$\alpha$ (5.64$\times$10$^{-11}$ erg~cm$^{-2}$~s$^{-1}$~arcsec$^{-2}$)
H$\beta$ (6.83$\times$10$^{-12}$), and 4364 
(3.18$\times$10$^{-13}$) lines, while the 5008 line reaches
the highest peak (6.71$\times$10$^{-12}$) at pixel 317  and a second relative 
maximum (5.14$\times$10$^{-12}$) at pixel 305
with a clear trough between these, having a relative minimum 
(3.58$\times$10$^{-12}$) occurring at pixel 310.
The F(H$\alpha$)/F(H$\beta$) ratio peaks at 8.269 also at pixel 311  
and hence the derived $c$(H$\beta$) reaches a peak value there of 2.08.

	Repeating the analysis for V72, we find that
the peak observed flux occurs in pixel 343 for
the H$\alpha$ (2.19$\times$10$^{-11}$) and 4364 (8.11$\times$10$^{-14}$) lines.
The H$\beta$ peaks in pixel 344 (3.19$\times$10$^{-12}$) and is lower at
pixel 343 (2.74$\times$10$^{-12}$). 
The 5008 line reaches
the highest peak (5.77$\times$10$^{-12}$) at pixel 350  and a second relative 
maximum (4.58$\times$10$^{-12}$) at pixel 336,
again with a distinct trough between these, having a relative minimum 
(2.85$\times$10$^{-12}$) occurring at pixel 344.
The F(H$\alpha$)/F(H$\beta$) ratio peaks at 7.994 at pixel 343  
where the derived $c$(H$\beta$) reaches a peak value of 2.01.
The fact that the peak surface brightnesses are much higher
for P159-350 in V2 compared with V72 is most likely because
the brightest part of the proplyd was better sampled and/or aligned
with the 1 pixel $\times$ 10 pixel (0.05$''$ by 0.5$''$) ``smallest rectangular aperture"
in the former visit than the latter.

\section{ELECTRON TEMPERATURE DETERMINATION}

\subsection{[\ion{O}{3}] Electron Temperature}

	The electron temperature $T_e$ is derived from the
intrinsic ratio I(5008)/I(4364)  using the following relation,
\begin{equation} 
T_e = 32966/[ln (I(5008)/I(4364)) - 1.701]~~  .   
\end{equation}
\noindent
Effective collision strengths are from Burke, Lennon \& Seaton (1989)
for $T_e$~= 10$^4$~K.
Transition probabilities (A-values) are from Froese Fischer \& Saha 
(1985).
Note that this holds in the low-$N_e$ limit, which should be valid
for Orion where $N_e$ values are less than the critical
densities ($N_{crit}$) for these lines.
The lowest $N_{crit}$ $\sim$6.4$\times$10$^5$ cm$^{-3}$ for the 5008 line,
which is well above $N_e$ values determined 
(e.g., Peimbert \& Torres-Peimbert 1977;
Osterbrock, Tran \& Veilleux 1992;
Esteban et~al.\ 1998).

	We continue with the analysis
using the tiles along the slit described above; the number of usable tiles
varies with the slit/visit and ranges from 79 to 96.
Equations (2), (3), and (5) are applied to the four emission lines to derive $T_e$.
Figure~2(a) shows the distribution of  $T_e$ versus position along slit~1.
For slit~1, position along the slit is relative to the location of peak surface 
brightness (S.B.) in the H$\alpha$ line, which occurs at P159-350.  
We set our zero corresponding to
coordinates, 
$\alpha$, $\delta$ = $5^{\rm h}35^{\rm m}15\fs94$, 
$-5^{\rm o}$23\arcmin50\farcs04
(Bob O'Dell, private communication). 
As mentioned in the last section, there is actually a 
relative minimum in the 
[\ion{O}{3}] 5008~\AA\ S.B. close to that position.
The lower curve in Figure~2(a) shows the observed 5008
S.B. (erg~cm$^{-2}$~s$^{-1}$~arcsec$^{-2}$),
which is displayed (unsmoothed) at the pixel level
providing 0.05$''$ spatial resolution along the slit.
The feature at the very left (SE) end of this slit is associated
with the prominent arc, part of which we used to define our position x2
and which was better observed in slit~2.
The open circles on the top ($T_e$) curve represent the individual 
tiles plotted at their midpoint.
The dashed lines are a linear interpolation across the tiles
that were deemed to have unreliable measurements because of proximity
to the fiducial bars.
There is a remarkably flat $T_e$ distribution with the notable
exception of an upward spike to 13600~K at the tile containing the bulk of
the proplyd H$\alpha$ emission.
For slit~1 and V2, the behavior is similar with a sharp
upward spike to 15050~K at the tile containing the bulk of
the proplyd H$\alpha$ emission.
It is most unlikely that these high temperatures are realistic 
(these tiles are omitted in the statistical analysis to follow)
and are the result of equation (5) not accounting for the
higher electron densities, probably in excess of 10$^6$~cm$^{-3}$,
in P159-350.  
Because the 5008 line will then suffer considerable collisional deexcitation,
$T_e$ derived using equation (5) will be overestimated
(e.g., Viegas \& Clegg 1994).
In the next section, we will evaluate the distribution of $T_e$ along slit~1,
as well as along the other slits,
in terms of $T_e$ variations.

	Figure~2(b) shows the distribution of  $T_e$ versus position along slit~2,
where position along the slit is relative to the location of the local
peak S.B. in the 5008 line that occurs at x2. 
Using the position of P159-350 and the specified offset from slit~1 to slit~2 
of $+$1.08s in RA and $-$8.15$''$ in Dec., we find x2 coordinates:
$\alpha$, $\delta$ = $5^{\rm h}35^{\rm m}16\fs96$, 
$-5^{\rm o}$23\arcmin57\farcs73.
The lower relative maximum near offset +14$''$ 
is where slit~2 passes through the ``downstream" tail of
P159-350.
Again there is a remarkably flat $T_e$ distribution.

	Figure~3(a) plots the distribution of  $T_e$ versus position along 
slit~4, where position along the slit is relative to the location of a feature
in the bar seen in the 6585 S.B.\ distribution (Fig. 3(c)) discussed in the 
next section.
Recall that this slit is aligned to point toward $\theta^1$~Ori~C. 
While there is an underlying decrease in the 5008 S.B. with increasing
distance from $\theta^1$~Ori~C, there is a leveling off (plateau) at 
the SE end of slit~4 of 5008 emission that remains substantial
well beyond the bar.  
This is consistent with the WFPC2 images of the region;
the composite ``drizzled" image by Walsh (1998) that includes filter F502N 
(which covers the 5008 line) depicted in green, 
shows a green hue in a parabolic shape that appears to be a wake emanating
from HH~203 and HH~204.
Walsh's image is available at
stecf.org/newsletter/webnews1/orion/m42col\_drizzle.jpg.
The point was also made in an earlier discussion of HH~203 and HH~204
(O'Dell et~al. 1997).
Along this slit, the $T_e$ distribution is also fairly flat,
especially in the region of higher S/N (see \S4.2).
There is a clear increase in the amplitude of the 
$T_e$ fluctuations for the tiles SE of the bar due
to poorer S/N in the 4364~\AA\ line.

	In Figure~3(b), we plot the distribution of  $T_e$ versus position along 
slit~5.  Here, the position along the slit is relative to the location of a feature
seen in the 6585 S.B.\ distribution (Fig. 3(d)) discussed below.
There is roughly a linear decrease in 5008 emission from the NW to the
SE end of slit~5 with a ``jump" occurring
between 18--22$''$ offset position.
As was the case  for slit~4, 
here too there remains substantial 5008 emission beyond the bar
that is evident in the green hue at this location in the Walsh image.  
The $T_e$ distribution here is flat similar to that for slit~4;
the increase in the $T_e$ amplitude range 
at the SE end of the slit is again 
due
to poorer S/N in the 4364~\AA\ line.

\subsection{[\ion{N}{2}] Electron Temperature} 

	The $T_e$ in the N$^+$ zone is derived from the
intrinsic ratio I(6585)/I(5756)
(see equations (2) and (4)).
Because the critical density for the 6585 line,
$\sim$7.7$\times$10$^4$ cm$^{-3}$ (at 10$^4$~K),
is substantially
less than $N_{crit}$ for the [\ion{O}{3}] 5008 line
($\sim$6.4$\times$10$^5$ cm$^{-3}$),
we do consider various $N_e$ values when deriving $T_e$. 
We also derive  temperatures utilizing two different
sets of \ion{N}{2} effective collision strengths~--
those calculated by Lennon \& Burke (1994) and 
by Stafford et~al. (1994).
We use the effective collision strengths for
10000~K;  these do not vary much with the $T_e$ range of interest in Orion.
The A-values used are discussed in Rubin et~al.\ 
(1998, Appendix A with original references therein).

	For non-zero densities, our derivation of $T_e$ is 
done in an iterative fashion, starting with an initial 
estimate in the low-$N_e$ limit.
Then the volume emissivities ($j$ values) 
for both the 5756 and 6585 lines are calculated
solving the statistical equilibrium equations for the six lowest energy
levels.   The $T_e$ value is then recomputed using the intrinsic 
ratio I(6585)/I(5756), $j$(6585), and $j$(5756).
The rapidly converging iteration is halted when $T_e$ changes by
less than 1~K.

	In this paper, we are interested in assessing the amount of
$t^2$ that can occur.  Thus, we limit our analysis here	by not
including a detailed study of density variations, which 
is the subject of future papers by the present authors.
Nevertheless, it is necessary to include some discussion
of $N_e$ with regard to deriving the 
[\ion{N}{2}] $T_e$.
As is well known, there is an inverse scaling of [\ion{N}{2}] $T_e$
with $N_e$.
First we perform the calculations for $T_e$ in the low-$N_e$ limit.
This provides a firm upper limit to the $T_e$ distributions evaluated for
the various slit/visit data.
We also repeat the full computations for 4 other $N_e$'s:
1000, 2000, 5000, and 10000~cm$^{-3}$.
The last value is more than high enough to bracket expectations for Orion,
with the exception of objects like the proplyds as will be
discussed below for P159-350.
The value of 5000 is our best single $N_e$ to cover the region of
slits 1 and 2 (e.g., Pogge, Owen \& Atwood 1992).
Their map of the electron density in the S$^+$ zone indicates that
$N_e$ is lower in the regions of our slit~4 and slit~5;
for these, we adopt a best single $N_e$~= 2000.
It is these respective densities that are used to depict
the distribution along slits 1, 2, 4, and 5 of 
[\ion{N}{2}] $T_e$ in Figure 2(c), 2(d), 3(c), and 3(d) respectively.
For the Figures, we have used the N$^+$ collision strengths of
Lennon \& Burke (1994) while results from this set, as well as
Stafford et~al. (1994) will be tabulated later.
The lower curve in each of these panels shows the 
6585 surface brightness distribution along the slit.

	Figure~2(c) shows the tiled [\ion{N}{2}] $T_e$ distribution 
versus position along slit~1 and is a positional match to Figure~2(a).
The lower curve here shows the observed 6585~\AA\ S.B.\ 
at the unsmoothed pixel level.
There again appears to be a relative minimum in the 
[\ion{N}{2}] 6585 S.B. at the centre pixel for P159-350
although it is not as well defined as the dip in the 5008 S.B. there.
As with the 5008 curve, the peak S.B. on the NW side of the proplyd
exceeds that on the SE side.  However, the overall 6585 emission
is much narrower in P159-350 than is the 5008 emission.
In totality, the structural behavior in the 6585 emission
compared with that in 5008 is consistent with the
N$^+$ region being more tightly confined than is the O$^{++}$ region
to the low-mass star of P159-350, which is expected if
the proplyd is externally ionized by $\theta^1$~Ori~C.
The prominent absolute maximum of the 6585 S.B.
is very close to the 1SW position, 20.26$''$ from P159-350,
which is somewhat blocked by the NW-side fiducial bar.
The narrow secondary peak at $\sim$13$''$ offset appears to
be associated with the much broader secondary maximum near 
offset $\sim$12$''$ in the 5008 S.B.\ of Figure 2(a).
There is perhaps an ionization boundary being observed here,
although the slit orientation is far from ideal to test this.
A slit pointing toward $\theta^1$~Ori~C would be better suited.
Once again, there is a notably flat $T_e$ distribution
with the exception of the proplyd.
In Figure 2(c), we have omitted the tile with the
highest derived $T_e$ of 20815~K.
The highest $T_e$ in the plot is at 13000~K.
Both these tiles are omitted from the subsequent statistical analysis.
These high derived $T_e$'s are undoubtedly pointing more to high $N_e$
values associated with the proplyd than high temperatures.
The $N_e$ assumed for the calculation of $T_e$ in Figure 3(c) is
5000~cm$^{-3}$, while $N_e$ for P159-350 may exceed 10$^6$~cm$^{-3}$.
Because the 6585 line emission will suffer enormously from collisional
deexcitation at more realistic densities, the derived $T_e$ would be much lower.
For slit~1, V2, there are two tiles at the proplyd
that have an enormous upward spike in $T_e$ to 29800 and 35850~K; 
both tiles are omitted from the statistical analysis in following sections.

	Figure~2(d) has the distribution of  [\ion{N}{2}] $T_e$ 
versus position along slit~2.
There is a fairly flat $T_e$ distribution with some
indication of a drop in $T_e$ SE of the lower fiducial bar.
The maximum 6585 S.B. in the lower curve
occurs near offset 28$''$.
This location is immediately adjacent to the 
narrow peak at $\sim$13$''$ offset in Figure 2(c), just mentioned.
The relatively small peak at x2 (offset 0) further supports the 
finding in the WFPC2 composite image that the emission arc which includes
the position x2 is mainly an [\ion{O}{3}] feature (see Figure~1). 
We note that x2 is part of the HH~529 complex and more specifically, 
the  easternmost feature called  170-358 
(see fig.~20 in Bally, O'Dell, \& McCaughrean 2000).  
According to their study, including the 
kinematics, 170-358  is an expanding bow shock seen from the side.

	Figure~3(c) plots the distribution of [\ion{N}{2}]  $T_e$ versus 
position along slit~4.
Position along this slit is relative to the location of the SE-most peak in 
S.B. in the 6585 line.
We set our zero corresponding to coordinates: 
$\alpha$, $\delta$ = $5^{\rm h}35^{\rm m}21\fs34$, 
$-5^{\rm o}$24\arcmin48\farcs21
determined from the 
position of our offset star
$\theta^2$~Ori~A (HD37041) with 
$\alpha$, $\delta$ = $5^{\rm h}35^{\rm m}22\fs9$, 
$-5^{\rm o}$24\arcmin57\farcs9.
There is a brighter feature that occurs where the slit crosses
HH~203 at offset near $-20''$.
Comparison with Figure 3(a) shows no enhancement in 5008.
This is consistent with the WFPC2 images of the region;
the composite image in Figure~1 
that includes filter F658N (isolating the 6585 line)
shows HH~203 as a reddish-orange object.
There is a very flat $T_e$ distribution even over the wide
range of S.B. and ionization structure along slit~4.

	In Figure~3(d), we plot the distribution of  $T_e$ versus position along 
slit~5.  
Here, the zero position along the slit is 
the SE-most 6585 S.B. peak of where there is a bifurcation in the bar
(see Figure~1).
This feature appears to mark the boundary
between the ionized region and the photodissociation region (PDR).
Using the position of $\theta^2$~Ori~A
and the specified offset from slit~4 to slit~5 
of $-1.87$s in RA and $-15.32''$ in Dec., we find:
$\alpha$, $\delta$ = $5^{\rm h}35^{\rm m}19\fs74$, 
$-5^{\rm o}$25\arcmin08\farcs14.
Beyond this position, the $T_e$ distribution becomes less reliable 
because of poorer S/N in the 5756~\AA\ line.
Overall, the $T_e$ is flat as has been the case throughout.

\section{FRACTIONAL MEAN-SQUARE TEMPERATURE VARIATIONS}

	Our STIS analysis above presents results in the plane of the sky.
The observations here do not address temperature fluctuation along the
line of sight, which may be characterized in terms of the average 
temperature $T_0$ and fractional mean-square $T_e$ variation ($t^2$) 
as defined by Peimbert (1967).  
\begin{equation} 
T_0~=~ {\int\ T_e\,N_e\,N_i\, dV
\over{\int\ N_e\,N_i\, dV}}~~,
\end{equation} 
\begin{equation} 
t^2~=~ {\int\ (T_e-T_0)^2\,N_e\,N_i\, dV
\over{T^2_0\int\ N_e\,N_i\, dV}}~~,
\end{equation} 
\noindent
where $N_i$ is the ion density $N(N^{+}$) or $N(O^{++}$).
The integration in equations (6) and (7)
is over the column defined by each
tile, and along the line-of-sight ($los$).
We are unable to measure the $t^2$ along the $los$ for any column
(cross section 1 tile).  
If there are $t^2$ along the $los$, we can say that
$T(4364/5008)$ $>$ $T_0$ 
or
$T(5756/6585)$ $>$ $T_0$ 
(e.g., Peimbert 1967; Rubin et~al.\ 1998).

	In the case of 	[\ion{O}{3}], for each tile, we have calculated $T(4364/5008)$.
Then the intrinsic flux I(5008), fully correcting F(5008) for extinction 
(see equ.\ 3), in each tile is used in conjunction with 
$T_e$~= $T(4364/5008)$ for that tile, 
and assumed constant along the $los$, to derive the following:
\begin{equation} 
I(5008) = K(5008) \int\ N_e\,N_i\,T_e^{-0.5}\,exp(-\chi/k\,T_e)\, dl~~
= K(5008, T_e) \int\ N_e\,N_i\, dl~~.
\end{equation} 
Here $\chi$ is the excitation energy above the ground state for the upper level 
of the 5008 transition, k is Boltzmann's constant, $K(5008)$ is known from
atomic data, and $K(5008, T_e)$ has finally incorporated the known $T_e$
factor with the atomic constants.
Here we again make the safe assumption of the low-N$_e$ limit 
(negligible collisional deexcitation) discussed earlier.

	For the case of [\ion{N}{2}], for each tile, we have calculated 
$T(5756/6585)$.
Then the intrinsic flux I(6585), fully correcting F(6585) for extinction 
(see equ.\ 4), in each tile is used in conjunction with 
$T_e$~= $T(5756/6585)$ for that tile and the chosen value for $N_e$, 
and assumed constant along the $los$, in the following relation:
\begin{equation} 
\int\ N_e\,N_i\, dl~~\propto ~{I(6585)\over{\epsilon(6585)}}~~.
\end{equation}

\noindent
Here 
$\epsilon(6585)$ is 
the normalized volume emissivity which is related to the usual volume emissivity
$\jmath(6585)$ by $\epsilon(6585)$~$\equiv$
$\jmath(6585)$/($N_e$$N_i$).
$\epsilon(6585)$ depends on the (fractional) population in a
given level obtained by solving the 6-level atom for the specific $T_e$ and $N_e$.  
It is not necessary to deal with the constant of proportionality for our 
purposes.

	Following Paper~I,
we define the average
$T_e$ ($T_{0,A}$) and fractional mean-square $T_e$ variation 
($t_A^2$) in the {\it plane of the sky}.

\begin{equation} 
T_{0,A}~=~ {\int \int\ T_e\,N_e\,N_i\, dl\, dA
\over{\int \int\ N_e\,N_i\, dl\, dA}}~~,
\end{equation} 

\begin{equation} 
t_A^2~=~ {\int \int\ (T_e-T_{0,A})^2\,N_e\,N_i\, dl\, dA
\over{T^2_{0,A}\int \int\ N_e\,N_i\, dl\, dA}}
= {\int \int\ T_e^2\,N_e\,N_i\, dl\, dA
\over{T^2_{0,A}\int \int\ N_e\,N_i\, dl\, dA}}\, -1~~,
\end{equation} 

\noindent
where $dA$ represents an element of surface area in the plane of the sky and 
the integration over $dl$ is for each tile along the $los$.
The proper weighting for each tile is provided by either equation (8)
or equation (9) depending on whether we are performing an 
[\ion{O}{3}] or [\ion{N}{2}] analysis.

	The results using the 
STIS data for the various slits and visits to determine $T_{0,A}$ and $t_A^2$
are summarized in Table~1.
These quantities for [\ion{O}{3}] are entered in the first column.
For slit~1, Visit~72 (see Figure 2(a)),  $T_{0,A}$~= 8258~K and $t_A^2$~= 
0.00925.  We have excluded the spike in $T_e$ at the proplyd.
The value for $t_A^2$ is remarkably low, although not as low as
the value of 0.0035 that we found for NGC~7009 (Paper~I).
The results from the other visits (V2 and V52) for slit~1 are close to
those for V72.
Our results for slit~2 and all three of the visits show $T_{0,A}$
values in close accord and similar to those for slit~1.
All these averages vary by less than 285~K.
By comparing $t_A^2$ for slits 1 and 2 in all 3 visits,
we see that all six numbers are small, ranging from 0.00682 to
0.0129.
For slits 4 and 5 the statistics are roughly comparable and
differ from those for slits 1 and 2 in the following ways:
$T_{0,A}$ is several hundred K smaller and $t_A^2$ is somewhat
higher, although the largest value, 0.0176, is still
notably small.
A simple visual comparison of the $T_e$ plots in Figs. 2(a), 
2(b), 3(a), and 3(b) 
leads one to conclude that $t_A^2$ looks larger for slits 4 and 5,
particularly with the larger $T_e$ excursions in the regions SE of the bar.
We note however that these data with poorer S/N have a minor effect on 
the statistics.
There is natural biasing against these tiles 
due to weaker emission in both 4364 and 5008
as can be seen from equation (8),
which in turn provides a smaller 
($\int N_e~N_i~dl$)--weighting 
of these tiles in  equations (10) and (11).

	All the remaining columns in Table~1 pertain to
the derivation of $T_{0,A}$ and $t_A^2$ for [\ion{N}{2}].
These are displayed for 5 different electron densities and
two different sets of effective collision strengths.
As described,  for slits 1 and 2, our preferred $N_e$ is 
5000~cm$^{-3}$, while for slits 4 and 5, it is 2000~cm$^{-3}$. 
We will now discuss the resulting numbers in these particular
density columns and also using the left-side column with the Lennon \& Burke
(1994) collision strengths, unless stated otherwise.
For slit~1, Visit~72 (see Figure 2(c)),  $T_{0,A}$~= 10226~K and $t_A^2$~= 
0.00695.  We exclude the high $T_e$ near P159-350 (see \S4.2).
The value for $t_A^2$ is even lower than found for this slit/visit
in the [\ion{O}{3}] analysis;
$T_{0,A}$ is nearly 2000~K higher.
There is close agreement among the N$^+$
$T_{0,A}$ and $t_A^2$ values for all the visits for slit~1;
the above intercomparison between N$^+$
and O$^{++}$ with the corresponding numbers
in the respective visits is also true.

	For slit~2, the $T_{0,A}$ numbers vary by more than
400~K between visits;  the respective $T_{0,A}$'s are smaller than
for slit~1 for V52 and V72 but slightly higher for V2.
All of the $t_A^2$ values are larger for slit~2 than for slit~1,
reaching 0.0146.

	For slits 4 and 5, the N$^+$ $T_{0,A}$
is several hundred to more than 1000~K 
smaller than for slits 1 and 2.
Compared with the O$^{++}$ $T_{0,A}$,
slits 4 and 5 are higher by 1343 and 1749~K, respectively. 
$t_A^2$ at 0.0175 is largest for slit~5
and nearly identical to the corresponding O$^{++}$ value, 
which is also the highest in the first column.
This larger $t_A^2$ may be attributed to the larger excursions
in $T_e$ in those tiles SE of the bar, where the S/N
in the 5756 \AA\ line is poorer.
A comparison of Figure 3(c) and 3(d) clearly indicates
a much lower 6585 S.B. SE of the bar for slit~5 than for slit~4.
Again we emphasize that these data with poorer S/N play only a minor
role in contributing to 
the integrals in  equations (10) and (11).
When there is weaker emission in both 5756 and 6585,
as can be seen from equation (9),
there is a smaller 
($\int N_e~N_i~dl$)--weighting 
of these tiles in  equations (10) and (11).

	The effect of substituting the Stafford et~al. (1994)
collision strengths is small for both slit~1 and slit~2
for all 3 visits: $T_{0,A}$ decreases by less than 200~K;
$t_A^2$ decreases by roughly 5--6 percent.	
The $T_{0,A}$ drop for slits 4 and 5 is slightly more,
up to a 276~K difference;
$t_A^2$ decreases by $\sim$6--7 percent.

	On the whole, the results presented in Table~1 
allow us to reach two robust conclusions.
First, $t_A^2$ is always very small ($\leq$ 0.0182)
for either the O$^{++}$ or N$^+$ analyses.
Second, the $T_{0,A}$ derived from the [\ion{O}{3}]  lines
is always significantly less than $T_{0,A}$ derived from the
[\ion{N}{2}]  lines.
These hold even when we consider 
for the N$^+$ analysis
a wide range in $N_e$
that should amply bracket the great bulk of the
plasma in Orion and also examine the influence
of using alternative collision strengths.

	In Appendix~A, we present an analysis of the uncertainties in
the derived $T_e$ values as well as the $t_A^2$ values.

\section{DISCUSSION AND CONCLUSIONS}

	We extend our earlier study of electron temperature
variations from the [\ion{O}{3}] (4364/5008) flux ratio
in the PN NGC~7009 (Paper~I)
to the Orion Nebula.
In addition, we use our STIS long-slit observations to
expand the analysis to measure 
$T_e$ variations from the [\ion{N}{2}] (5756/6585) flux ratio.
The observations here do not address $T_e$ fluctuation along the line of sight
through the specific O$^{++}$ region or, likewise, through the N$^+$ region.
We analyse both the [\ion{O}{3}]  and [\ion{N}{2}] 
data sets to derive the average $T_e$ and fractional mean-square 
$T_e$ variations in the plane-of-the-sky, which we call  $T_{0,A}$ and $t_A^2$.
We assume for each square column (projection of 1 STIS tile 0.5$''$ square 
on the plane-of-the-sky) that the 
plasma along the line of sight is isothermal at the $T(4364/5008)$
in the case of O$^{++}$ or 
$T(5756/6585)$ in the case of N$^+$.
For the latter case, we consider a large range of $N_e$ values,
which should be sufficient for Orion, in order to produce Table~1.
The analysis for each $N_e$ assumes that it is constant for the 
entire STIS slit length and throughout the sheet projected 
through the nebula along the $los$.

	Fluctuations in $T_e$ (and $N_e$) along the $los$ are inevitable.
We can make some comments about how our results for $t_A^2$ 
might be adjusted by $T_e$ variations along the $los$.
The relationship between $T(4364/5008)$ and $T_0$ for the O$^{++}$ region is

\begin{equation} 
T(4364/5008) = T_0~[ 1 + 0.5 (91200/T_0 - 3) t^2]~~,
\end{equation} 
and 
between $T(5756/6585)$ and $T_0$ for the N$^+$ region is

\begin{equation} 
T(5756/6585) = T_0~[ 1 + 0.5 (69000/T_0 - 3) t^2]~~,
\end{equation} 
(e.g., Peimbert 1967; Rubin 1969).
With the $T(4364/5008)$-- or $T(5756/6585)$--values
for Orion (see Figures 2 and 3) or indeed for
\ion{H}{2} regions in general,
$T_0$ will be smaller than these temperatures inferred from the
forbidden line flux ratios here.
Rather than repeat some examples of how the differences
depend on $t^2$, the reader is referred to our comments
in the earlier paper (\S7 in Paper~I).

	We do not have the data here
to characterize $T_e$ variations in 3-dimensions (3-D). 
It is useful to define an {\it overall} 3-D average $T_e$ ($T_{0,V}$)
and fractional mean-square $T_e$ variation ($t_V^2$).
These single values apply for the {\it entire source}.
Equations (6) and (7) define these specific values when the integration
is over the entire volume.  
We note that for a spatially unresolved object
(total integrated fluxes observed in the aperture), 
{\it not} the case for Orion, 
$t_V^2$~= $t^2$ and a calculation of $t_A^2$ is meaningless.

	Our measurements of $T_e$ reported here are an average
along each line of sight.
Because each element of area treated in the plane of the sky
represents a column which has already created a spatially averaged
temperature along the $los$
(e.g., see fig.\ 1 in Rubin 1969), it is likely that 
the value for $t_V^2$ is substantially higher than $t_A^2$.
Measurements of $t^2$ along various sight lines 
appear to be the most direct way to reliably gauge $t_V^2$.
Therefore, despite finding remarkably low $t_A^2$, we cannot completely
rule out much larger temperature fluctuations along the $los$.
Further work, beyond the scope of this paper, is underway that
will use modeling as well as additional observational data
in an effort to better determine the relationship between
$t^2$, $t_A^2$, and $t_V^2$.

	Finding [\ion{N}{2}] temperatures
that are higher than [\ion{O}{3}] temperatures is not a new result.
For Orion, this has been known for years (e.g.,
Baldwin et al.\ 1991). 
What is new/significant here is that we have results
from many more sight lines, more than 700 represented by
the individual tiles, with
improved spatial resolution (0.5 arcsec squares) and
excellent spatial registration between the 
[\ion{N}{2}] and [\ion{O}{3}] data sets that HST affords.
Of the roughly 700 tiles,
about half represent independent sight lines, with the other
half having spatial overlap due to the repeated visits for slits~1 and 2.

	The fact that the [\ion{N}{2}] temperatures
are higher is also expected on theoretical grounds
and again, not something novel that we have uncovered.
We enumerate three factors that contribute to the
predicted inequality.
First, the cooling of the [\ion{O}{3}] 
5008, 4960 \AA\ lines in the O$^{++}$ region 
is uniquely efficient for nebular 
conditions and elemental abundances that
prevail in Galactic \ion{H}{2} regions, including Orion. 
On the other hand, in the ``singly ionized" region
(where the dominant O and N ions are O$^+$, N$^+$)
there is no coolant that is nearly as efficient as
[\ion{O}{3}] 5008, 4960 is in the O$^{++}$ zone. 
Additionally for Orion, the blister geometry
likely results in higher average electron densities
in the singly ionized region compared with
the O$^{++}$ region because the former is closer to the PDR.
The higher $N_e$ in the N$^+$ region would also contribute to
less efficient cooling.

	Second, there is the predicted hardening
of the stellar ionizing photons at progressively
larger distances from the exciting star.
This is due to the functional form of the H photoionization
(predominantly) cross section, which diminishes steeply 
with higher frequency from its value near threshold.
At larger distances from the exciting source,
the average energy per photoionization will increase and
thus the heating rate will also increase.  This causes a rise in
$T_e$ with an increase in distance (other factors being equal).
This has been known for many years (e.g., Rubin 1968).

	Third, there is a possibility that there is a 
contribution to the production of [\ion{N}{2}] 5756~\AA\ emission
(and to a lesser extent 6585) by recombination and cascading.
Under some conditions, this may provide significant routes into 
the upper energy level of the 5756 transition 
that are not negligible compared with collisional processes.
This was examined by Rubin (1986).
If there were a significant ``recombination" contribution to
the observed emission, we would have overestimated the 
$T_e$ derived from the [\ion{N}{2}] 5756/6585 flux ratio.
With our two independent, detailed models for the Orion Nebula
(Baldwin et~al.\ 1991 and Rubin et~al.\ 1991),
the recombination contribution appears to be small.
We discussed this in a subsequent paper where these
models were ``retrofitted" with more in-common input parameters,
including the Stafford et~al.\ (1994) N$^+$ collision
strengths, to allow closer intercomparisons 
(see end of \S4 in Rubin et~al.\ 1998).

	All three of the above effects are included in our 
photoionization, plasma simulation models just mentioned.
The predicted [\ion{N}{2}] and [\ion{O}{3}] temperatures 
for the entire volume are 8721~K and 7704~K in the Rubin et~al.\ (1991)
model and 8649~K  and 7692~K in the retrofit 
(Rubin et~al.\ 1998) using Stafford et~al.\ cross sections.
When the Lennon \& Burke (1994) N$^+$ collision strengths
are used instead of the Stafford et~al.\ set,
the respective temperatures become 8706~K and 7701~K.

	Finally, we encapsulate the findings here for the proplyd P159-350.
We find large local extinction as evidenced by the dramatic increase in
the observed F(H$\alpha$)/F(H$\beta$) ratio along slit~1.
For V2, this ratio peaks at 8.27 which implies a 
$c$(H$\beta$) of 2.08 from the 	Balmer decrement. 
Similar values are found for our V72 observations:
F(H$\alpha$)/F(H$\beta$) peaks sharply at the proplyd position
reaching a ratio of 7.99,
where the derived $c$(H$\beta$)~= 2.01.
Because the adjacent nebular values for 
$c$(H$\beta$) are much lower, the extinction must be associated
with the P159-350 environs.

	A comparison of the 
[\ion{O}{3}] 5008 
and [\ion{N}{2}] 6585
surface brightnesses in the vicinity of P159-350 
(see Figures 2(a) and 2(c)) shows 
for both lines that the S.B. on the NW side of the proplyd
exceeds that on the SE side.  
However, the 6585 emission is much narrower than is the 5008 emission
which provides evidence of ionization stratification consistent with the
N$^+$ region being more tightly confined than is the O$^{++}$ region
to the low-mass star of P159-350.
This ``inverse \ion{H}{2} region" behavior 
is precisely what is expected due to 
the external ionization source $\theta^1$~Ori~C.

	The derived $T_e$ distributions were all notably flat 
except at the position of the proplyd observed in slit~1.
The very high derived $T_e$'s at the location of P159-350
are no doubt due more to high $N_e$
values associated with the proplyd than high temperatures.
As stated, the calculations used here to assess $T_e$ have not 
accounted for $N_e$ values likely to exceed 10$^6$~cm$^{-3}$ in P159-350.
Because both the 5008 line and even more so the 6585 line emission 
will suffer substantial collisional deexcitation at such high densities, 
the derived $T_e$ for both [\ion{O}{3}] and [\ion{N}{2}] 
would be much lower.

	Each of the above mentioned facets of
P159-350 deserves further study.
The smallest spatial resolution element was 
1 pixel (0.05$''$) in the spatial direction by 10 pixels (0.5$''$) in
the dispersion direction.
It would be highly desirable to obtain higher spatial resolution of P159-350, 
which can be readily achieved by observing with a narrower STIS slit.
The narrower slit would enable better spectral resolution as well
and permit a detailed mapping of P159-350 in the
\ion{C}{3}] 1907, 1909 \AA\ lines.
The ratio of their fluxes is an excellent diagnostic of $N_e$
even at the high $N_e$ values expected in proplyds.
Improved spectral resolution in the optical lines 
will also help sort out kinematics from spatial structure,
which was a difficulty with the 0.5$''$ slit width we used.

\acknowledgments

This paper is based on observations made with the NASA/ESA {\it Hubble Space
Telescope}, obtained at the Space Telescope Science Institute (ST~ScI).
Support for proposal \#GO-7514 was provided by NASA through a grant from 
ST~ScI. ST~ScI is operated by the Association of Universities for Research in 
Astronomy, Inc., under NASA contract NAS5-26555.  
Valuable contributions were made by Naman Bhatt, Ephrat Bitton, Brent Buckalew,
Bob O'Dell, Chris Ortiz, Aaron Svoboda, and Abby Wong.
We thank Bob O'Dell and Janet Simpson for reading the paper and providing useful comments.
We much appreciate the exemplary service of our ST~ScI contact scientist Paul Goudfrooij
and program coordinator Ray Lucas.
RHR acknowledges support from the Long-Term Space Astrophysics (LTSA)
program, NASA/Ames Research Centre contract NCC2-9018 with Orion
Enterprises, and thanks Scott McNealy for providing a Sun workstation.

\clearpage

\appendix

\section{Error Analysis}

The values found for $t_A^2$ are very small.  It is important to
determine whether even these small fluctuations in $T_e$ from tile to
tile are a real signal or the result of measurement uncertainty.

 From equation (5) for O$^{++}$, it can be shown that the mean square
uncertainty in temperature, expressed as $t_e^2$, has the following
dependence on the uncertainties $\delta I_n$ and $\delta I_a$ of the
nebular ($I_n$) and auroral ($I_a$) line fluxes, respectively:

\begin{equation}
t_{e}^2 = \left( \delta T_{e} / T_{e} \right) ^2 = \frac{T_{e}^2}{{c_1}^2}
\left[ \left(\frac{\delta I_a}{I_a}\right)^2 +
\left( \frac{\delta I_n}{I_n}\right)^2 \right],
\label{firsteq}
\end{equation}
where $c_1=32966$.  We use this as an approximation for N$^{+}$ too,
with $c_1=24933$. 

The uncertainty $\delta I$ is principally the measurement error in the
line flux for a tile.  The S/N for the flux of the auroral lines
([\ion{O}{3}] 4364 and [\ion{N}{2}] 5756 here) is much lower than that
for the nebular lines ([\ion{O}{3}] 5008 and [\ion{N}{2}] 6585 here),
and so the uncertainty of the weaker auroral line will dominate the
uncertainty $\delta T_{e}$.  The uncertainty $\delta I$ also depends
on the (differential) extinction correction, but in the present case
this is a small effect compared to the uncertainty in the observed
auroral line fluxes (hence, $I/\delta I \sim F/\delta F =$ S/N ).

The three independent methods used to calculate the S/N of the auroral
lines are discussed in the following sections.  We use $T_{e} \sim
T_{0,A}$ to calculate the corresponding $t_e^2$.

\subsection{Method 1}

The flux integrated over the line profile as found using the ``blkavg
sum'' technique in IRAF is
\begin{equation}
F^{line} =  \sum_{i=1}^m D F_{i}^{line+cont} - \frac{m}{n}\sum_{j=1}^n 
D F_{j}^{cont},
\end{equation}
where $F_{i}$ is the monochromatic flux for the tile (erg cm$^{-2}$
s$^{-1}$ \AA$^{-1}$), $D$ is the dispersion conversion (\AA/pixel),
$n$ is the number of pixels in the spectrum used to define the
$continuum$ and $m$ is the number of pixels used for the $line+cont$.
Since the auroral line is weak, the rms fluctuation in the $line+cont$
can be assumed to be the same as in the $continuum$, $\delta
F_j^{cont}$.  Thus, the rms of the line flux is
\begin{equation}
\delta F^{line} = D \sqrt{m} ( \delta F_j^{cont} ) \sqrt{1+m/n}.
\end{equation}
In our measurements we used $m=n$.  The above three equations are used
to create the entries for method~1 (M1) in Table~\ref{appendixtable}.
Values are given for both the [\ion{O}{3}] 4364 and [\ion{N}{2}] 5756
lines for all slits (for slits with multiple visits, a representative
case is given).

\subsection{Method 2}

In this method, the line profiles from the STIS spectra are fit with a
template (a Gaussian convolved with a slit of width 0.5\arcsec) using
a combined Gauss-Newton and modified Newton algorithm (Numerical
Algorithms Group (NAG) - E04FDF) to find the minimum least-squares
solution and the corresponding variances of the variables used in the
fit (NAG-E04YCF).  The slit width is held constant, but due to
anamorphic magnification in the dispersion direction, the plate scale
differs from line to line which results in different slit widths
expressed as pixels (see Bowers \& Baum 1998).

The variables used in the template fitting are the flux $F^{line}$,
the FWHM and central wavelength of the Gaussian, and the slope and
mean brightness of the continuum.  The variances reflect how well the
model fits the data.  Keeping all of these as variables gives one a
conservative estimate of the variance of each.  Note that a poor
template would lead to an overestimate of the error.  The templates
were tested on the strong nebular lines and generally fit very well.
However, there are instances where the surface brightness changes
across the slit (in the dispersion direction), so that the implicit
assumption of uniform illumination is not perfect.  Therefore, the
estimated uncertainty from method 2 is an upper limit.  As is seen in
Table~\ref{appendixtable} it agrees well with the error estimated
using method 1.

\subsection{Method 3}

This method utilizes the multiple sets of observations for slit~1 and
slit~2 in the three visits V2, V52 and V72 and examines the
reproducibility of the line fluxes and $T_e$.  There are adequate
fiducial sources visible in all three visits so that the spectra may
be shifted in the spatial direction to achieve alignment to $\sim$1
pixel.  The proplyd P159-350 in slit 1 is particularly useful for this
and serves for slit~2 as well because the relative positional HST
offsets between slit~1 and slit~2 are not subject to the acquisition
uncertainty.  There is also positional alignment uncertainty in the
dispersion direction which is more difficult to estimate.  For all
three visits, P159-350 was seen in the 0.5$''$ wide slit; although it
was blocked for the most part by the East fiducial bar in V52, some of
the emission still peeks out both ends.  We noted at the end of
section~3 that P159-350 has a higher peak surface brightness in V2
than in V72.  We estimate that the positional alignment of the tiles
perpendicular to the spatial direction between the three visits is
within 0.2$''$ so that there is substantial area overlap (perhaps
60\%) of our tiles between the three visits.  Comparison of the
nebular line strengths from visit to visit shows excellent
reproducibility ($< 3\%$ per tile for the O$^{++}$ line).  This is
much smaller than the auroral line differences found among successive
visits.  The measurement error of the weaker auroral lines is of
course higher.  Furthermore lack of registration is a more serious
issue for the auroral lines (from a much higher energy level) if there
are real $T_{e}$ fluctuations on this spatial scale; in the limit of
no overlap, one would expect $t_{e}^2 \sim t_{A}^2$ even if there were
no measurement errors.  Therefore, having $t_{e}^2$ from method~3
larger than from M1 and M2 is not unexpected.

For each tile in common between V2, V52 and V72, both an average
value ($F_a$) and the standard deviation ($\delta F_a$) of any single
determination of $F_a$ were calculated.  The representative value
entered as M3a in Table~\ref{appendixtable} is the median $\delta
F_a/F_a$ for each of the slits; as in methods 1 and 2, the $\delta
T_e$ listed is what is implied by this S/N, through equation
\ref{firsteq}.

For each in-common tile, we also calculated $T_e$ for each visit, then
the average $T_e$ and the actual standard deviation ($\delta T_{e}$)
of any single determination of $T_e$.  The median $\delta T_{e}$ for
each slit is included in Table~\ref{appendixtable} as M3b.  From these
median values we calculated the tabulated $t_e^2$.  Methods 3a and 3b
agree closely, as they should given the good reproducibility of the
nebular lines
(i.e., the effect of $\delta F_n$ in equation
\ref{firsteq} is negligible).

The analysis summarized in Table~\ref{appendixtable} shows that the
measured $t_{A}^2$ is a real signal not dominated by the noise and/or
measurement errors.  The two exceptions are for the O$^{++}$
measurements for slits 4 and 5 which include tiles beyond the Orion
bar (SE part of the slit) where the surface brightness for lines of
that ion, and hence the S/N, is low.  Accordingly, we repeated the
entire analysis for only the NW half of these slits.  The results
recorded in Table~\ref{appendixtable} show a lower $t_A^2$, as
anticipated, and a lower $t_e^2$ as well which is a smaller fraction
of $t_A^2$.

Where there is a contribution of measurement error to the apparent
$t_A^2$, the actual $t_A^2$ is smaller, by about $t_e^2$.  This
reinforces our conclusion that $t_A^2$ is very small.

\clearpage 

\begin{deluxetable}{llrllrllrllrllr}
\tabletypesize{\scriptsize}
\tablecaption{Values of $T_{0,A}$ and $t^2_A$ \label{tsquared}}
\tablewidth{0pt}
\tablehead{\colhead{O$^{++}$} & \multicolumn{14}{c}{N$^+$}\\
\cline{2-15}\\
\colhead{$N_e$ (cm$^{-3}$) \tablenotemark{a}} & \multicolumn{2}{c}{0} &&
\multicolumn{2}{c}{1000} && \multicolumn{2}{c}{2000} &&
\multicolumn{2}{c}{5000} && \multicolumn{2}{c}{10000}\\
\cline{2-3} \cline{5-6} \cline{8-9} \cline{11-12} \cline{14-15}}
\startdata

\sidehead{Slit 1, Visit 2}
8151\tablenotemark{b} & 11104 & (10605)\tablenotemark{c} && 10881 &
(10458) && 10670 &  (10317) && 10114 &  (9929) &&  9381 &  (9386)\\
1.04\tablenotemark{b} & 0.860 & (0.795)                  && 0.844 &
(0.785) && 0.831 & (0.777) && 0.791 & (0.752) && 0.730 & (0.712)\\

\sidehead{Slit 1, Visit 52}
8232  & 11236 & (10727) &&  11010 & (10578) && 10796 & (10435) && 10232
& (10042) &&  9488 &  (9491)  \\
0.887 & 0.647 & (0.592) &&  0.633 & (0.584) && 0.620 & (0.576) && 0.584
& (0.553) && 0.531 & (0.517)  \\

\sidehead{Slit 1, Visit 72}
8258  & 11232 & (10723) &&  11005 & (10573) && 10791 & (10429) && 10226
& (10036) &&  9481 &  (9485)  \\
0.925 & 0.765 & (0.703) &&  0.750 & (0.694) && 0.736 & (0.685) && 0.695
& (0.659) && 0.636 & (0.620)  \\

\sidehead{Slit 2, Visit 2}
8142  & 11171 & (10665) &&  10945 & (10517) && 10732 &  (10373) && 10170
&  (9982) &&  9429 &  (9433)  \\
1.15  & 1.05  & (0.956) &&  1.02 & (0.942) && 1.00 & (0.929) && 0.941
& (0.890) && 0.853 & (0.831)  \\

\sidehead{Slit 2, Visit 52}
8074  & 10675 & (10208) &&  10463 & (10068) && 10262 &  (9933) &&  9733
&  (9564) &&  9038 &  (9046)  \\
1.29  & 1.60  &  (1.48) &&   1.57 &  (1.46) &&  1.54 &  (1.44) &&  1.46
&  (1.39) &&  1.33 &  (1.30)  \\

\sidehead{Slit 2, Visit 72}
8358  & 10909 & (10424) &&  10691 & (10280) && 10484 &  (10141) &&  9938
&  (9760) &&  9221 &  (9227)  \\
0.682 & 1.43  &  (1.31) &&   1.40 &  (1.30) &&  1.37 &  (1.28) && 1.30
&  (1.23) && 1.18  &  (1.15)  \\

\sidehead{Slit 4, Visit 5}
7790  &  9473 &  (9107) &&   9298 &  (8991) &&  9133 &  (8879) &&  8698
&  (8572) &&  8125 &  (8143)  \\
1.57  & 0.985 & (0.918) &&  0.968 & (0.906) && 0.952 & (0.896) && 0.906
& (0.867) && 0.837 & (0.821)  \\

\sidehead{Slit 5, Visit 5}
7678  &  9789 &  (9394) &&   9603 &  (9271) &&  9427 &  (9151) &&  8962
&  (8825) &&  8353 &  (8369)  \\
1.76  & 1.82  &  (1.69) &&   1.79 &  (1.66) &&  1.75 &  (1.64) &&  1.66
&  (1.58) &&  1.52  & (1.49)  \\

\enddata
\tablenotetext{a}{electron density assumed for N$^+$ emission-line
region; $0$ denotes the low density limit}
\tablenotetext{b}{upper row $T_{0,A}$ (K); lower $100\ t^2_A$}
\tablenotetext{c}{using effective collision strengths from Lennon \& Burke 1994
(Stafford et al.\ 1994)}
\end{deluxetable}

\clearpage

\begin{deluxetable}{ccccccccc}
\footnotesize
\tablecaption{Weak line uncertainties and the associated fractional mean-square temperature variation, $t_e^2$}
\tablehead{
\colhead{Line} &
\colhead{Slit} & \colhead{Visit} & \colhead{Method\tablenotemark{a}} &
\colhead{$(\delta F_a/F_a)_{median}$} &
\colhead{$T_{0,A}$}  & \colhead{$\delta T_e$} & \colhead{$100 t^2_e$} &
\colhead{$100 t^2_A$}
}
\startdata
4364 & 1 & V72 & M1 & 0.1434 & 8258 & 297 & 0.129 & 0.925 \\
 & &  & M2 & 0.1132 & & 234 & 0.080 & \\
 & & \tablenotemark{b} & M3a & 0.1826 & & 378 & 0.209 & \\
 & & \tablenotemark{b} & M3b & & & 388 & 0.221 & \\
4364 & 2 & V72 & M1 & 0.2003 & 8358 & 424 & 0.258 & 0.682 \\
 & & & M2 & 0.2577 & & 546 & 0.427 & \\
 & & \tablenotemark{b} & M3a & 0.2593 & & 550 & 0.432 & \\
 & & \tablenotemark{b} & M3b & & & 513 & 0.377 & \\
4364 & 4 & V5 & M1 & 0.3129 & 7790 & 576 & 0.547 & 1.57 \\
 & &  & M2 & 0.5350 & & 985 & 1.60 & \\
4364 & 4NW & V5 & M1 & 0.2435 & 7665 & 434 & 0.320 & 0.973 \\
 & &  & M2 & 0.1550 & & 276 & 0.130 & \\
4364 & 5 & V5 & M1 & 0.4274 & 7678 & 764 & 0.991 & 1.76 \\
 & &  & M2 & 0.4527 & & 810 & 1.11 & \\
4364 & 5NW & V5 & M1 & 0.3235 & 7683 & 579 & 0.568 & 1.42 \\
 & &  & M2 & 0.3310 & & 593 & 0.595 & \\

5756 & 1 & V72 & M1 & 0.0499 & 10226\tablenotemark{c} & 209 & 0.042 & 0.695\tablenotemark{c} \\
  & &     & M2 & 0.0616 &         & 258 & 0.064 &       \\
  & &  \tablenotemark{b} & M3a & 0.0937 &        & 393 & 0.148 & \\
  & &  \tablenotemark{b} & M3b & & & 388 & 0.144 & \\
5756 & 2 & V72 & M1 & 0.1188 & 9938\tablenotemark{c} & 471 & 0.224 & 1.30\tablenotemark{c} \\
  & &     & M2 & 0.1090 &        & 432 & 0.189 &      \\
  & &  \tablenotemark{b}  & M3a & 0.1879 &        & 744 & 0.561 &      \\
 & & \tablenotemark{b} & M3b & & & 729 & 0.538 & \\
5756 & 4 & V5 & M1 & 0.1491 & 9133\tablenotemark{d} & 499 & 0.298 & 0.952\tablenotemark{d} \\
  & &    & M2 & 0.1226 &         & 410 & 0.202 &       \\

5756 & 5 & V5 & M1 & 0.1459 & 9427\tablenotemark{d} & 520 & 0.304 & 1.75\tablenotemark{d}  \\
  & &    & M2 & 0.1284 &         & 458 & 0.236 &       \\

\enddata
\tablenotetext{a}{Methods 1-3 (M1-M3) are discussed in the text.}
\tablenotetext{b}{V2,V52 and V72 are used for Method 3.}
\tablenotetext{c}{$N_e=5000$ cm$^{-3}$ using effective collision strengths from Lennon \& Burke 1994}
\tablenotetext{d}{$N_e=2000$ cm$^{-3}$ using effective collision strengths from Lennon \& Burke 1994}
\label{appendixtable}
\end{deluxetable}

\clearpage

\begin{figure}
\plotone{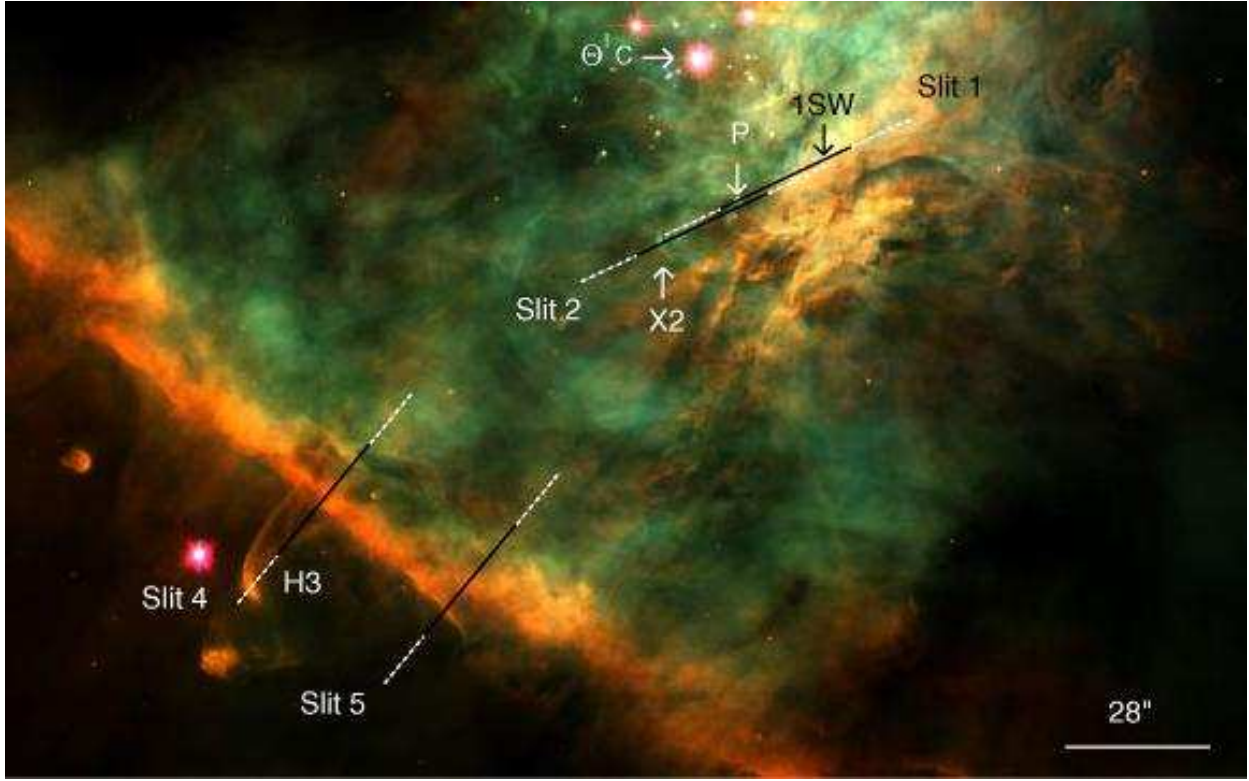}
\caption{This shows the position of the four STIS slits overlaid
on a composite WFPC2 image with 
[\ion{O}{3}] (5008~\AA ) blue,
H$\alpha$ green, and 
[\ion{N}{2}] (6585~\AA ) red.
The CCD observations presented here are taken with the
STIS long-slit 52$''$ $\times$ 0.5$''$.
The middle portion of each slit shown in black 
represents the smaller length (28$''$) for which there are other data
obtained with the MAMA detectors.
Abbreviated labels stand for positions described in the text:
P for the proplyd P159-350; H3 for HH~203.
Note that the star $\theta^2$~Ori~A is the unlabeled red feature just above the 4 in Slit~4.
N is up and E to the left.}
\end{figure}

\clearpage

\begin{figure}
\plotone{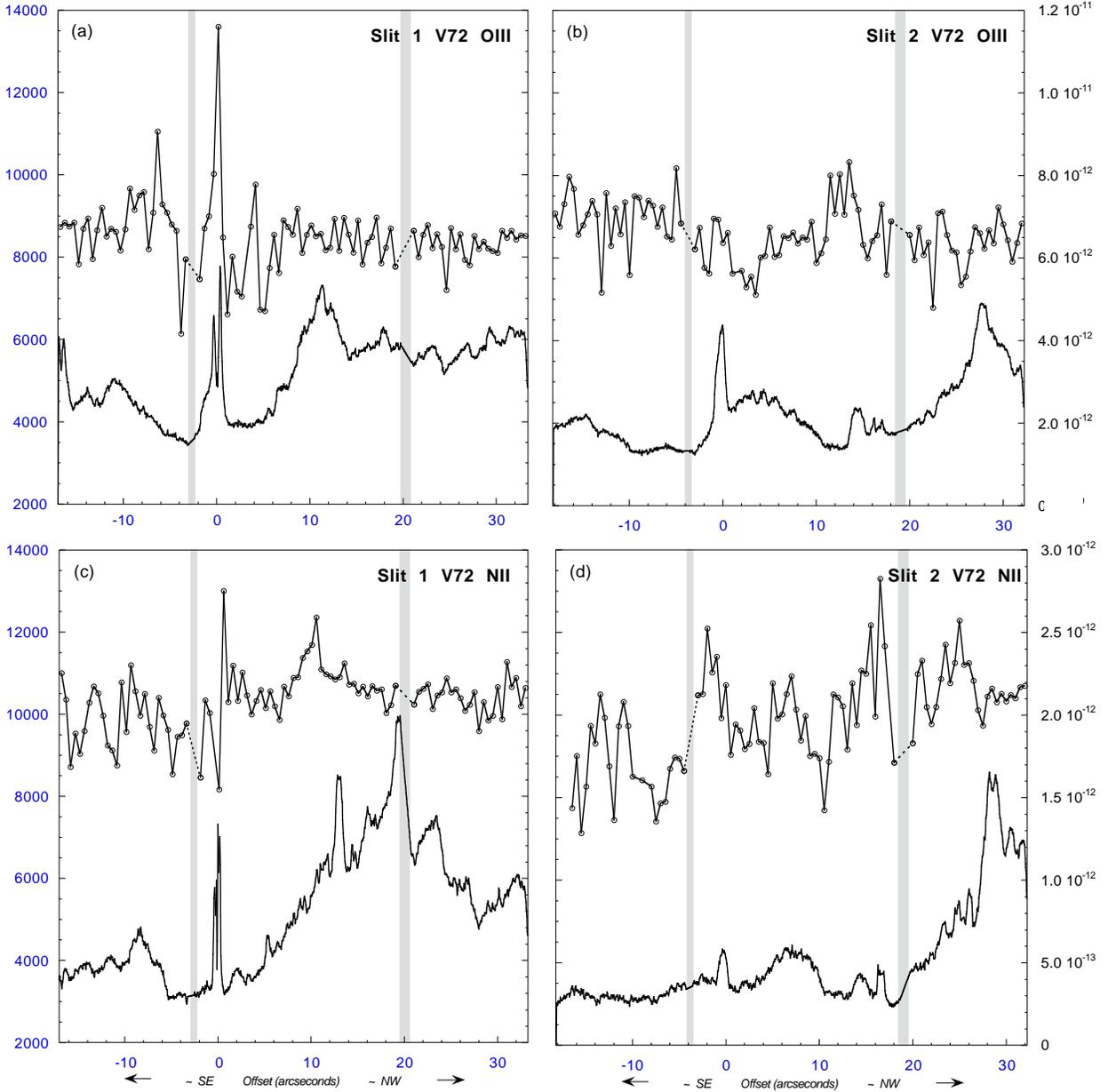}
\caption{{\bf (a)} Slit~1, V72: Plot of $T_e$ determined from the [\ion{O}{3}] 4364/5008 flux ratio
versus position along the STIS long-slit.  The analysis is in terms of
tiles  that are 0.5$''$ square (matching the slit width).  
The open circles represent the individual tiles plotted at their midpoint.
The dashed straight lines are interpolations across the two fiducial bars,
indicated by the gray area, where data are unreliable.
Positional measurement along the slit is from $\sim$SE to $\sim$NW (see 
Figure~1).  The zero point is described in the text.
The bottom curve shows the observed [\ion{O}{3}] 5008~\AA\
surface brightness in units of erg~cm$^{-2}$~s$^{-1}$~arcsec$^{-2}$
displayed unsmoothed at the pixel level.
~~~~~~~~{\bf (b)} same as (a) for Slit~2, V72.\break 
{\bf (c)} same as (a) except the upper curve is a
plot of $T_e$ determined from the [\ion{N}{2}] 5756/6585 flux ratio
assuming $N_e$~= 5000~cm$^{-3}$;
the lower curve shows the observed [\ion{N}{2}] 6585~\AA\
surface brightness.
~~~~~~~~{\bf (d)} same as (c) for Slit~2, V72.}
\end{figure}

\clearpage

\begin{figure}
\plotone{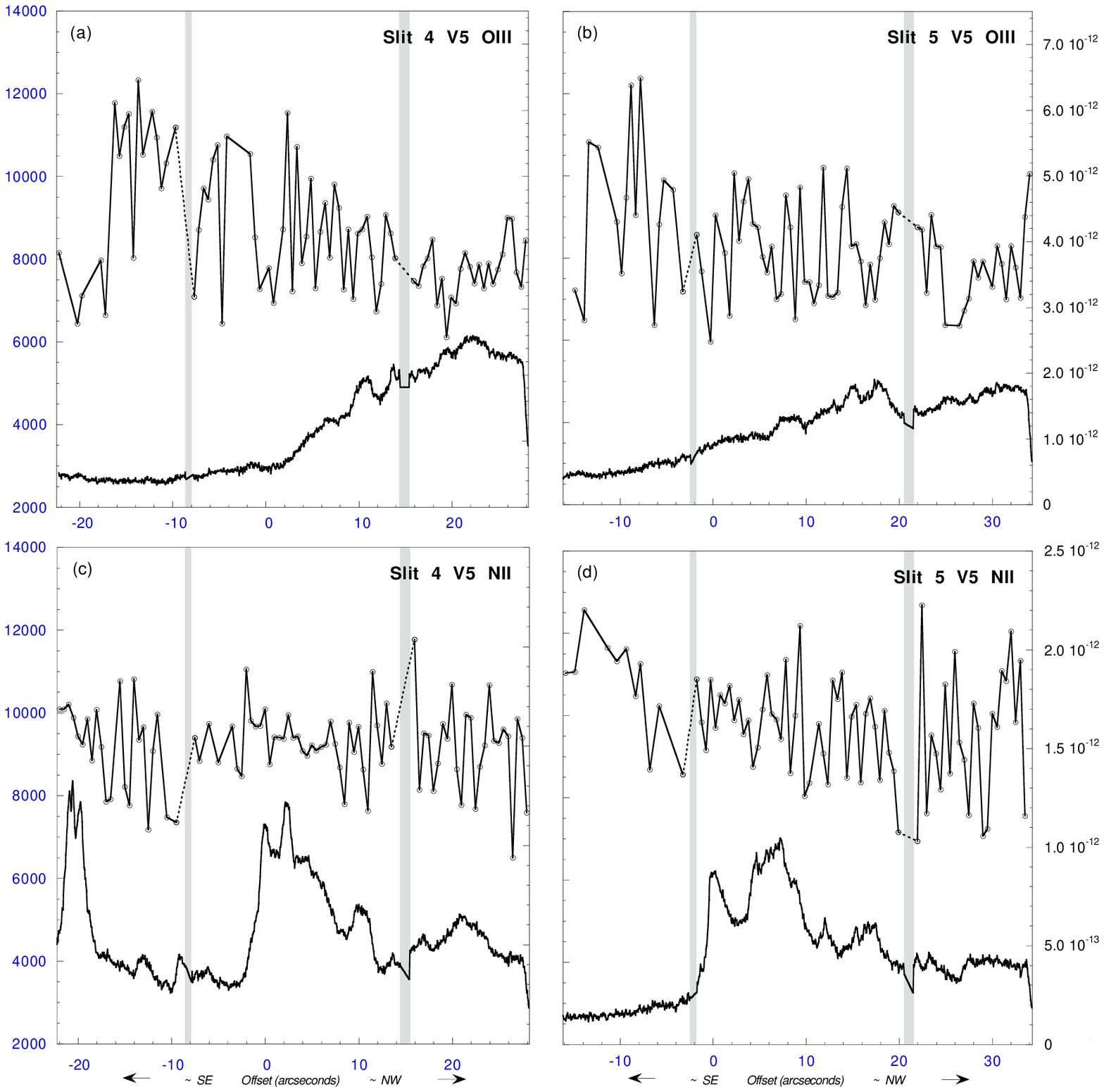}
\caption{{\bf (a)} Slit~4, V5: Same as Fig.2 (a)
~~~~~~~~{\bf (b)} same as (a) for Slit~5, V5.\break
{\bf (c)} Slit~4, V5: Same as Fig.2 (c) except that $N_e$~= 2000~cm$^{-3}$
is assumed.\break
{\bf (d)} same as (c) for Slit~5, V5.}
\end{figure}


\clearpage


\begin{thebibliography}{}

\bibitem[]{}
Baldwin J.A., Ferland G.J., Martin P.G., Corbin M., 
Cota S., Peterson B.M., Slettebak A., 1991, ApJ, 374, 580

\bibitem[]{}
Bally J., O'Dell C.R., McCaughrean M.J., 2000, AJ, 119, 2919

\bibitem[]{}
Bohlin R., Collins  N., Gonnella A., 1998, Instrument Science Report,
STIS 97-14, (Baltimore: STScI) 

\bibitem[]{}
Bohlin R., Hartig G., 1998, Instrument Science Report,
STIS 98-20, (Baltimore: STScI) 

\bibitem[]{}
Bowers C., Baum S., 1998, Instrument Science Report, STIS 98-23,
(Baltimore: STScI) 

\bibitem[]{}
Burke V.M., Lennon D.J., Seaton, M.J., 1989, MNRAS, 236, 353

\bibitem[]{}
Esteban C., Peimbert M., Torres-Peimbert S., Escalante V., 1998, MNRAS, 295, 401

\bibitem[]{}
Esteban C., Peimbert M., Torres-Peimbert S., Garcia-Rojas J., 
Rodriguez M., 1999, ApJS, 120, 113


\bibitem[]{}
Froese Fischer C., Saha H.P., 1985, Phys.\ Scr., 32, 181

\bibitem[]{}
Grevesse N., Sauval A.J., 1998, in Space Science Reviews, 85, 161


\bibitem[]{}
Harrington J.P., Seaton M.J., Adams S., Lutz J.H.,
 1982, MNRAS, 199, 517

\bibitem[]{}
Kingdon J.B., Ferland G.J., 1998, ApJ, 506, 323


\bibitem[]{} 
Kwitter K.B., Henry R.B.C., 1998, ApJ, 493, 247

\bibitem[]{} 
Leitherer C., et al., 2001, STIS Instrument Handbook, Version 5.1, (Baltimore: STScI)

\bibitem[]{} 
Lennon D.J., Burke V.M., 1994,  A\&AS, 103, 273

\bibitem[]{}
Liu X.-W., 2002, in ASP Conf.\ Ser.,
Planetary Nebulae: Their Evolution and Role in the Universe,
Astron.\ Soc.\ Pac., San Francisco (submitted)


\bibitem[]{}
Liu X.-W., Luo S.-G., Barlow M.J., Danziger I.J., Storey P.J.,
2001, \mnras, 327, 141 

\bibitem[]{}
Liu X.-W.,  Storey P.J., Barlow M.J., Clegg R.E.S., 1995, \mnras, 272, 369

\bibitem[]{}
Liu X.-W., Storey P.J., Barlow M.J., Danziger I.J., Cohen M., Bryce M., 
2000, \mnras, 312, 585


\bibitem[]{}
Luo S.-G., Liu X.-W., Barlow M.J.,  2001, \mnras, 326, 1049

\bibitem[]{}
Martin P.G., Rubin R.H., Ferland G.J.,  Dufour R.J., O'Dell C.R., 
Baldwin J.A., Hester J.J., Walter D.K., 
1996, BAAS, 28, 1416

\bibitem[]{}
O'Dell C.R., Hartigan P., Lane W.M., Wong S.K., Burton M.G., Raymond J.,
Axon D.J., 1997, AJ, 114, 730


\bibitem[]{}
O'Dell C.R., Wen Z.,  1994, ApJ, 436, 194

\bibitem[]{}
Osterbrock D.E., Tran H.D., Veilleux S.,  1992, ApJ, 389, 305  

\bibitem[Peimbert 1967]{pei67} 
Peimbert M., 1967, \apj, 150, 825

\bibitem[]{}
Peimbert M., Costero R., 1969, Bol.\ Obs.\ Ton.\ y Tacu., 5, 3

\bibitem[]{}
Peimbert M., Torres-Peimbert S., 1977, MNRAS, 179, 217 

\bibitem[P\'equignot et al.\ 2002]{peq01} 
P\'equignot D., et al., 2002, in Proceedings
Ionized Gaseous Nebulae.
RevMexAA (Serie de Conferencias), 12, 142
eds. W.J. Henney, J. Franco, M. Martos, M. Pena 

\bibitem[]{}
Pogge R.W., Owen J.M.,  Atwood B., 1992, ApJ, 399, 147

\bibitem[]{}
Rubin R.H., 1968, ApJ, 153, 761

\bibitem[]{}
Rubin R.H., 1969, \apj, 155, 841

\bibitem[]{}
Rubin R.H., 1986, \apj, 309, 334 

\bibitem[]{}
Rubin R.H., Bhatt N.J., Dufour R.J., Buckalew B.A.,  Barlow M.J.,
Liu X.-W., Storey P.J., Balick B., Ferland G.J., Harrington J.P.,
Martin P.G., 2002, \mnras, 334, 777 (Paper~I)


\bibitem[]{}
Rubin R.H.,  Dufour R.J., Ferland G.J., Martin  P.G., O'Dell C.R., 
     Baldwin J.A., Hester J.J., Walter D.K., Wen Z., 1997,  
     \apj, 474, L131

\bibitem[]{}
Rubin R.H., Martin  P.G., Dufour  R.J., Ferland G.J.,
Baldwin J.A., Hester J.J., Walter D.K., 1998,  ApJ, 495, 891

\bibitem[]{}
Rubin R.H., Simpson J.P., Haas M.R., Erickson E.F., 1991, ApJ, 374, 564


\bibitem[]{}
Stafford R.P., Bell K.L., Hibbert A., Wijesundera W.P., 1994, 
    \mnras, 268, 816


\bibitem[Storey \& Hummer 1995]{sto95}
Storey P.J., Hummer D.G., 1995, \mnras, 272, 41 

\bibitem[]{}
Tsamis Y.G., Barlow M.J., Liu X.-W., Danziger I.J., Storey P.J., 2002,
\mnras ~(in press)

\bibitem[]{}
Viegas S.M., Clegg R.E.S., 1994, MNRAS, 271, 993

\bibitem[]{}
Walsh J.R., 1998 ST-ECF Newsletter, number 25 (see Cover) 
\end{thebibliography}
\end{document}